\documentclass[twocolumn]{aastex631}

\usepackage{graphicx}
\usepackage{multirow}

\begin{document}

\title{Uncovering the Next Galactic Supernova with the Vera C.~Rubin Observatory}

\author[0000-0003-0776-8859]{John Banovetz}
\affiliation{Lawrence Berkeley National Laboratory, 1 Cyclotron Road, Berkeley, CA 94720, USA}
\affiliation{Physics Department, Brookhaven National Laboratory, Upton, NY 11973, USA}

\author[0000-0002-7397-2690]{Claire-Alice H\'ebert}
\affiliation{Physics Department, Brookhaven National Laboratory, Upton, NY 11973, USA}

\author[0000-0002-5209-872X]{Peter B.~Denton}
\affiliation{Physics Department, Brookhaven National Laboratory, Upton, NY 11973, USA}

\author[0000-0002-4934-5849]{Dan Scolnic}
\affiliation{Department of Physics, Duke University Durham, NC 27708, USA}

\author[0000-0002-8713-3695]{An\v{z}e Slosar}
\affiliation{Physics Department, Brookhaven National Laboratory, Upton, NY 11973, USA}

\author[0000-0003-2035-2380]{Chris Walter}
\affiliation{Department of Physics, Duke University Durham, NC 27708, USA}

\begin{abstract}
Supernovae are observed to occur approximately 1-2 times per century in a galaxy like the Milky Way. Based on historical records, however, the last core-collapse galactic supernova observed by humans occurred almost 1,000 years ago. Luckily, we are well positioned to catch the next one with the advent of new neutrino detectors and astronomical observatories. Neutrino observatories can provide unprecedented triggers for a galactic supernova event as they are likely to see a supernova neutrino signal anywhere from minutes to days before the shock breakout causes the supernova to brighten in optical wavelengths. Given its large etendue, the Vera C.~Rubin Observatory is ideally positioned to rapidly localize the optical counterpart based on the neutrino trigger. In this paper we simulate events to study the efficiency with which supernovae are optimally localized by the Vera C.~Rubin Observatory. We find that the observatory is ideal for initial localization of nearly all observable supernova triggers and has a 57-97\% chance of catching any supernova based on theoretical stellar mass density predictions and observations. We provide an analysis of optimal filter selection and exposure times and discuss observational caveats.
\end{abstract}


\section{Introduction} \label{sec:intro}
Core-collapse Supernovae (CCSNe), the explosions of massive stars ($>8 M_{\odot}$) \citep{Smartt2009}, are one of the most influential and energetic transient events in our Universe. CCSNe are responsible for the creation of neutron stars and black holes, as well as many of the heavy elements necessary for life on Earth \citep[e.g.][]{Burbidge1957}, and influence galactic evolution \citep[e.g.][]{Dekel2019}. They are also one of the most common transient events in the night sky, with thousands of SNe 
% accessible to our instrumentation happening 
observed each year \citep{Aleo2023}. Despite the influence and regularity of these events, as nearly all of the CCSNe in recent history have been detected as unresolved sources outside the Milky Way, with the exception of SN 1987A, many mysteries remain such as the level of explosion/ejecta asymmetry and the role of a companion star. 

While the astronomical community has learned a considerable amount from extragalactic CCSNe, a galactic CCSN will be immensely valuable, providing unique insight into the explosion mechanism, the surrounding environment that creates a CCSN, and neutrino properties. The last galactic CCSN observed and recorded from multiple sources occurred over 1,000 years ago \citep{Mayall1939}\footnote{While Cas A was more recent, there is some ambiguity to the whether or not it was observed \citep[see][]{Thorstensen2001}}. While there are some SN remnants that are younger than 1,000 years \citep[e.g.][]{Carlton2011}, the majority of these are located in the heart of the Milky Way where dust would have extincted the light out of range visible to human eyes. Recent studies on CCSN rates for a galaxy similar to the Milky Way place estimate there should be 1-2 CCSNe every century \citep{Li2011}.

Obstruction by dust and the galactic bulge may explain the historic lack of observed galactic SN. However, in the recent era of multimessenger astronomy, we will be able to overcome these limitations via neutrino detectors. The death of a massive star begins when it runs out of fuel for fusion and is left with an iron core that ultimately collapses in on itself, forming either a black hole or a neutron star depending on the progenitor mass \citep{Sukhbold2016}. If a CCSN forms, the infalling material will bounce back from the proto-neutron star, which is at nuclear densities, creating an explosion of neutrinos. While all SNe create some neutrinos, CCSNe convert roughly 99\% of the energy of the explosion into neutrinos \citep[see][for review]{Mueller2019}. These will travel to Earth unimpeded by dust and inform astronomers of a galactic CCSN event. The electro-magnetic (EM) counterpart of the explosion is generated after the neutrinos, when the shock breaks out of the stellar envelope.  The exact time between the release of neutrinos and photons depends on the radius and stellar type of the progenitor \citep{Kistler2013}. Estimates of the delay between receiving a neutrino signal and detection of the light from shock break-out (SBO) will therefore provide useful information. 

We are entering an unprecedented age of astrophysical neutrino detection. Sensitive neutrino detectors such as Super-K \citep{Fukuda2003}, IceCube \citep{1612.05093}, and JUNO \citep{1507.05613} are already running, and more detectors such as DUNE \citep{2002.03005} and Hyper-K \citep{1805.04163} are expected to come online over the next decade. All of these can, or will be able to, detect neutrinos from a galactic CCSN. 
Some of these detectors even provide a precise constraint of the probable sky area; for example, Super-K can locate a source at 10 kpc to a pointing accuracy within 3-7 degrees \citep{Machado2022,Kashiwagi2024}. If there is a galactic CCSN, a combination of these detectors will be sending out an alert via the SNEWS network \citep{SNEWS2021}.
While neutrino detectors provide invaluable information, localizing and detecting the electro-magnetic counterpart can be especially challenging depending on where the next galactic CCSN is located in our galaxy.

The Vera C.~Rubin Observatory is an ideal facility to detect the EM counterpart from a neutrino trigger. The Rubin optical system employs six different filters\footnote{Only five filters can be loaded into the filter exchange system at once, and thus be available on short notice.} able to cover wavelengths of 350--1050\,nm and has a field of view of almost 10 square degrees, roughly the size of Super-K's trigger radius \citep{LSST,Walter2019}. Combining these with a large 6\,m effective-area mirror means that Rubin is uniquely suited to quickly cover the search area with relatively deep exposures, possibly mitigating dust extinction by observing with a redder band. Fast slew speed is another one of the greatest advantages of using Rubin for follow-up of multi-messenger galactic CCSN alerts; Rubin can potentially detect the SBO within a few minutes of receiving an alert. Using Rubin, we can ensure that astronomers will be able to quickly locate the next galactic CCSN even if it is on the other side of the Milky Way. 

As such, it would benefit the astronomical community to create an observing strategy in the event there is a galactic CCSN during the lifetime of Rubin's ten year Legacy Survey of Space and Time (LSST). While there are models that suggest that there may be a gravitational wave detection before a star goes CCSN \citep[e.g.][]{Choi2024}, but there has never been such a detection and so we will be focusing on neutrino alerts from a galactic supernova for our study. While there is also additional neutrino information that may come in the form of presupernova neutrinos from the silicon burning phase hours to days earlier, this is not the trigger that this study focuses on. This flux is extremely dim in comparison and is only detectable for the closest CCSN within $\lesssim1$ kpc which comprises only a few dozen likely progenitors, but may provide additional information for a galactic CCSN follow up campaign \citep{Odrzywolek2004,Simpson2019,Mukhopadhyay2020,Abusleme2024}. \cite{Duverne2025} also discusses optical follow-up of neutrino alert from a galactic supernova. While their focus was on triangulation from neutrino detectors and following up with Rubin Observatory and/or TAROT \citep{Klotz2008}, we assume that there will be pointing information from Super-K.

Following \cite{Andreoni2024}, the general idea is that when a neutrino trigger occurs and the alert with the search area is sent, Rubin will slew to the search area and begin taking images as soon as possible in order to capture the SBO, which can appear minutes to days after the neutrino trigger depending on the progenitor type. Rubin should repeatedly tile the search area in order to find objects with rapidly changing brightness. Once an object is found that fits the rising luminosity of a CCSN, Rubin should issue an alert and continue to take observations in order to provide valuable cross-calibrations with other observatories.

In this paper we discuss possible observing strategies utilizing Rubin to find the electro-magnetic counterpart of a galactic CCSN during the time of LSST, building upon previous papers \citep{Adams2013,Walter2019}. 
Section \ref{sec:Models} discusses the various models and simulations we used to make our predictions about the next galactic CCSN. We discuss the results of combining these models in Section \ref{sec:results}. In Section \ref{sec:Discussion}, we describe the observing strategy given the values described in Section \ref{sec:results}. And finally, Section \ref{sec:conclusion} concludes our study.

\section{Models and Simulations} \label{sec:Models}

\begin{figure}[!thp]
    \centering
    \includegraphics[width=\columnwidth]{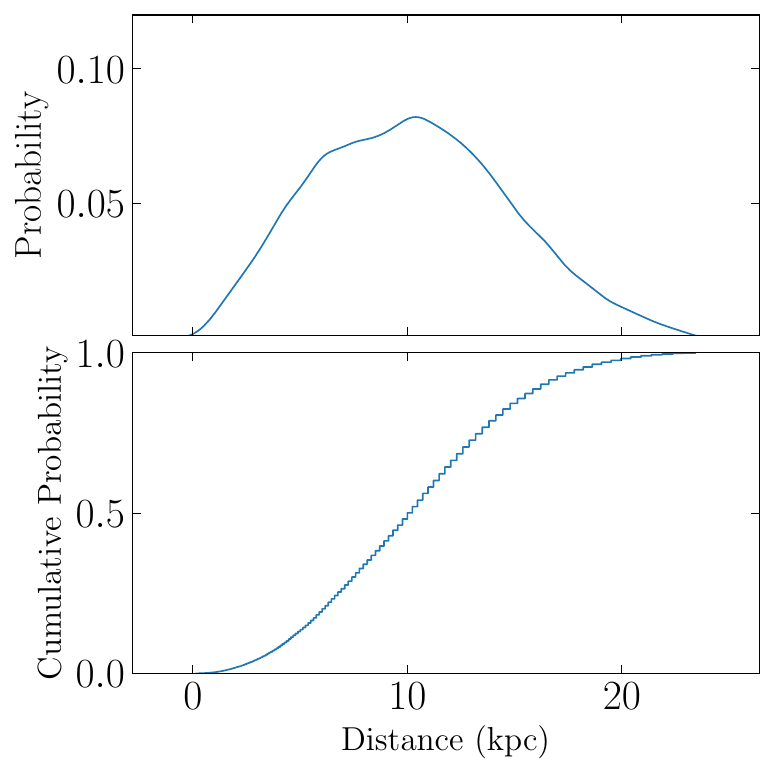}
    \caption{\textbf{Top:} Probability distribution of the distance away from Earth for the TRILEGAL CCSN candidates. \textbf{Bottom:} Same as the panel above as a cumulative distribution. }
    \label{fig:dist_prob}
\end{figure}

\begin{figure}[!thp]
    \centering
    \vspace*{-1cm}
    \includegraphics[width=0.49\textwidth]{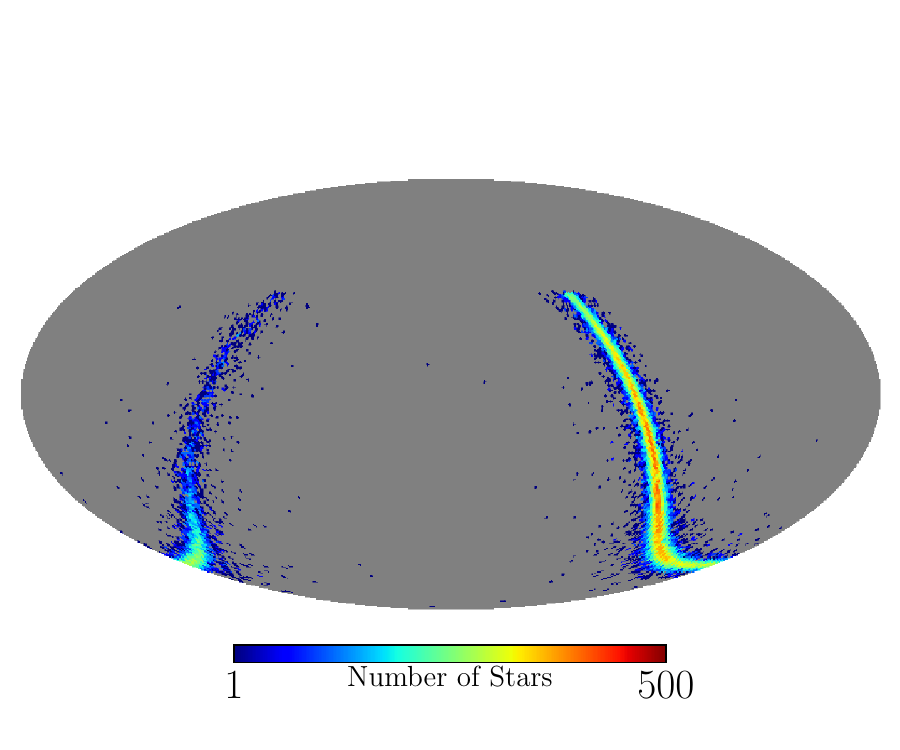} \\
    \includegraphics[width=0.49\textwidth]{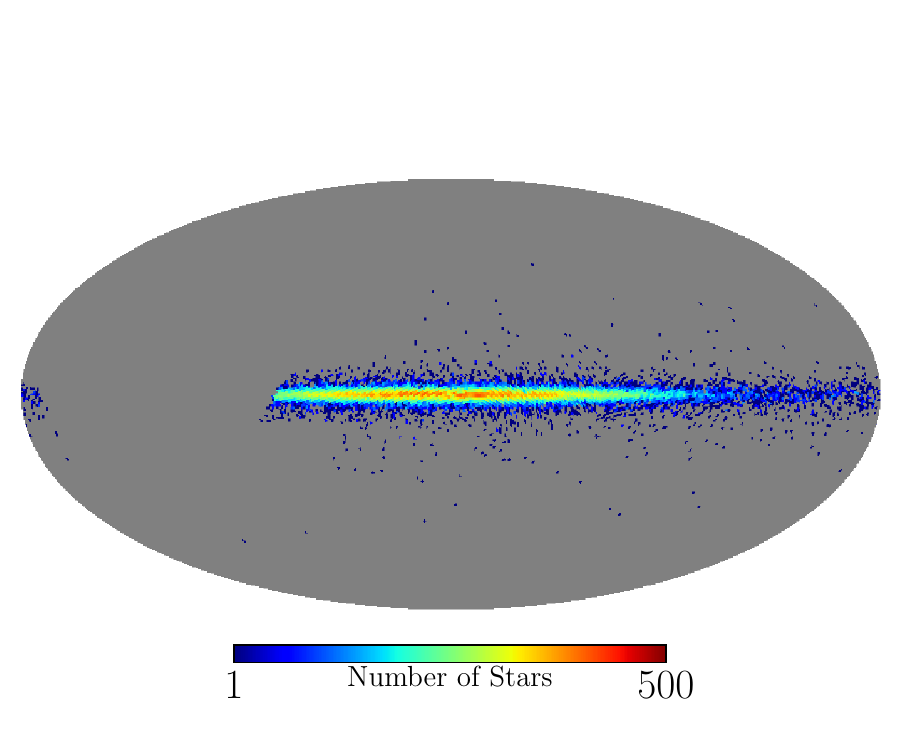}
    \caption{\textbf{Top:} The distribution of TRILEGAL stars in a Molleview Celestial projection. \textbf{Bottom:} Same as the above figure but in galactic coordinates. Both of these show the visibility cutoff due to the location of Rubin.}
    \label{fig:star_distribution}
\end{figure}

\begin{figure}[!htp]
    \centering
    \includegraphics[width=0.49\textwidth]{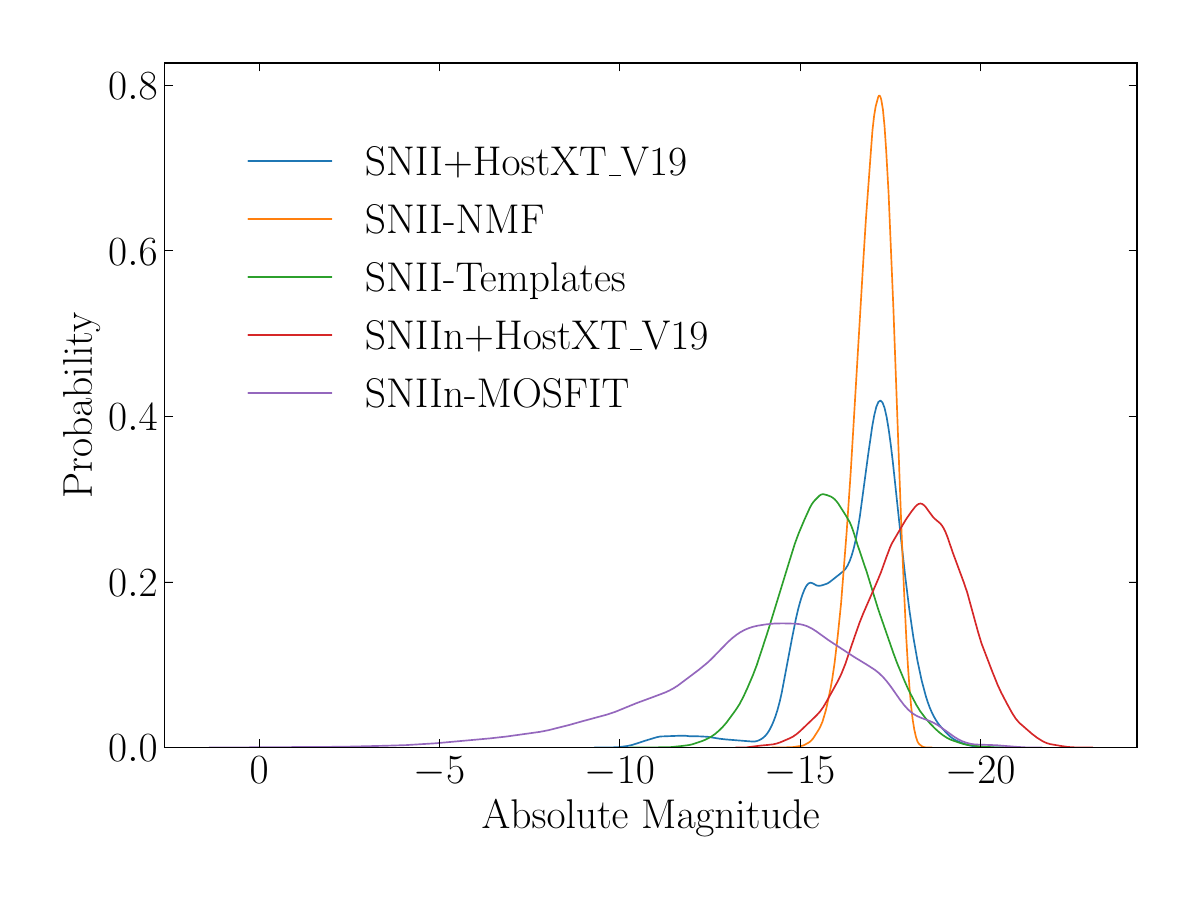}
    \caption{Probability distribution of events and their corresponding peak absolute LSST r-band magnitude for Type II CCSN SCOTCH Models. For our study, we focused on the SN-II Template model.}
    \label{fig:Scotch_Abs}
\end{figure}

In order to predict whether Rubin can detect the next galactic CCSN we need a model of the Milky Way with CCSN candidates at various distances and extinctions. For our model of the the Milky Way, we use existing data created from the TRILEGAL \citep{Girardi2005,Girardi2012,Marigo2017} code, a catalog of simulated stellar sources from the Milky Way that would be observable as a part of the LSST imaging campaign (henceforth LSST TRILEGAL) \citep{TRILEGAL}\footnote{\url{https://datalab.noirlab.edu/query.php?name=lsst_sim.simdr2}}.
We make several cuts to the existing dataset to produce our CCSN candidate sample. The first is to use only the single star evolution dataset provided by LSST TRILEGAL containing 10.6 billion stars. While there is evidence, from both observations \citep[e.g.][]{Drout2023} and simulations \citep{Zapartas2019,Muller2019}, that many CCSN progenitor systems are binary systems, we choose the single star catalog for simplicity. 
% and in order to probe more of the Milky Way: the single star evolution dataset contains 10.6 billion stars compared to the 1.61 billion binary systems in the binary evolution dataset.
Second, we restrict our candidates to stars in the Milky Way (LSST TRILEGAL includes stars in the Magellanic Clouds). 
Neutrinos are detectable from CCSNe in the Magellanic Clouds---famously, in the case of 1987A, the neutrino detection was made in retrospect, after the electro-magnetic counterpart was discovered \citep{Hirata1987}. 
At distances of 50\,kpc and 62\,kpc for the LMC and SMC, respectively, the number of the SN neutrinos reaching the Earth will be considerably lower than from a galactic SN, greatly reducing the resulting pointing angle resolution. 
Third, we only choose stars that have the potential to create a CCSN ($>10 M_\odot$ for a zero-age main-sequence mass) \citep{Smartt2009}. We choose this cutoff mass to consider only iron core-collapse and avoid electron-capture SNe, which could potentially have different core-collapse energetics and neutrino physics \citep[see][for review]{Wang2026}\footnote{Tests of our results with a $8 M_\odot$ cutoff show that the differences in our results are of the order of $1\%$ and are thus negligible.}.
After these selections, the number of potential CCSN candidates reduces to 39,723 from the 10.6 billion stars available from TRILEGAL. This is smaller than expected, given the initial mass functions used in TRILEGAL \citep{Chabrier2001}, but still within recent estimates \citep[see][]{Kroupa2019} especially given the upper mass limit of 60 solar masses.
The distance distribution of the candidates can be seen in Figure \ref{fig:dist_prob}, and projections of the locations onto the sky can be seen in Figure \ref{fig:star_distribution}.

There are many factors that determine the specific CCSN type of a progenitor \citep[see][ for review]{Gilkis2025}; only one of these, its zero-age main-sequence mass, is available in LSST TRILGAL for our candidates. As such, we assign to each of CCSN candidates a CCSN type following the assumption that 70\% of these stars will result in a Type II SN and 30\% will result in Type Ib/c \citep{Li2011}. We use the Simulated Catalogue of Optical Transients and Correlated Hosts (SCOTCH) \citep{Lokken2023}, which provides a large number of CCSN models, to create the light curves for all the candidates.
Figure \ref{fig:Scotch_Abs} shows example distributions of peak absolute magnitude for the Type II CCSN models in SCOTCH, highlighting the range of values possible.
For this paper, we assume that all Type II explosions will follow the SN-II Templates model (shown in green in Figure \ref{fig:Scotch_Abs}), and all Type Ib/c will follow the SNIb-Templates model.
To model these explosions in the Milky Way, we apply extinction to each explosion model based on the distance modulus and sky position provided by LSST TRILEGAL for each candidate.  Though LSST TRILEGAL also provides $A_{V}$ values, we wanted to test higher values than its maximum of $A_{V}\sim 30$. Instead, we use the dust map from \cite{SFD98} (henceforth SFD), with the corrections detailed in \cite{Schlafly2011}, and apply the extinction using the \texttt{extinction}\footnote{http://github.com/kbarbary/extinction} package. To get the SFD $A_{V}$ values (which reach $A_{V}>100$) at the candidate locations we used \texttt{dustmaps} \citep{dustmaps} and assumed an R$_{V}$ of 3.1. 

\section{Results} \label{sec:results}
\begin{figure*}[!htp]
    \centering
    \includegraphics[width=0.99\textwidth]{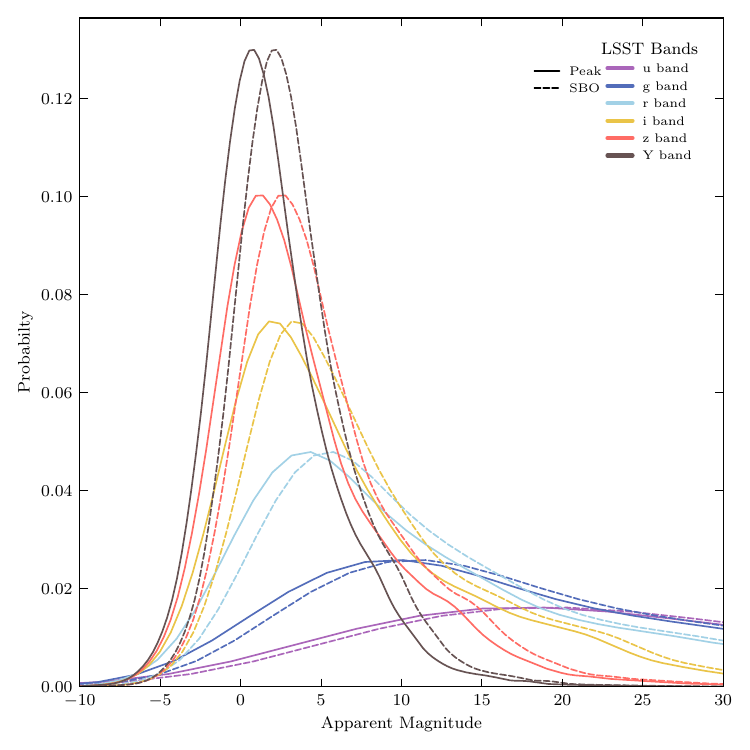}
    \caption{The probability distribution of the peak (solid) and SBO (dashed) apparent magnitudes for all LSST TRILEGAL CCSN candidates with CCSN rates applied, using the SCOTCH SN-II and SN-Ib Template models used to represent Type-II and Type-I, respectively. }
    \label{fig:PDF}
\end{figure*}

\begin{figure*}[!htp]
    \centering
    \includegraphics[width=0.99\textwidth]{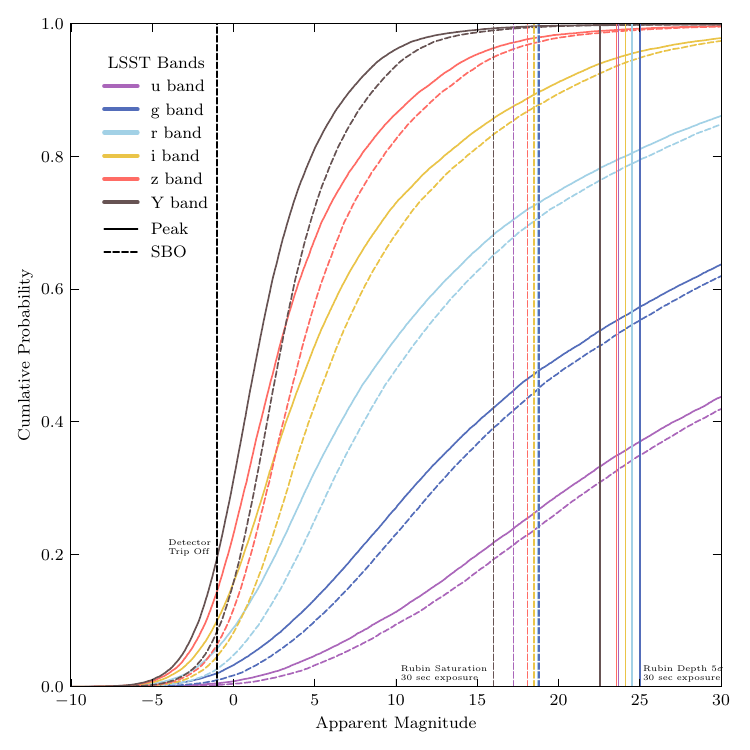}
    \caption{Cumulative distribution of values shown in Figure \ref{fig:PDF}. The Rubin magnitude depth limits and saturation limits of a 30 second exposure at an airmass of 1 are shown in solid and dashed lines, respectively. The black dashed line represents the magnitude of a CCSN that could trip off a detector. To match this to the numbers in Table \ref{tab:Sat_sig}, the intersection of the `Saturation' line and the curve corresponds to the `Saturated' column, the difference between the $5\sigma$ exposure depth intersection and the saturation intersection corresponds to the `Visible' column, and the `Too Faint' column is 1 minus the $5\sigma$ intersection.}
    \label{fig:ECDF}
\end{figure*}

\begin{table*}[!tp]
    \centering
    \begin{tabular}{|c||c|c|c||c|c|c||c|}
       \hline
       \multirow{2}{*}{Band} & \multicolumn{3}{c||}{Peak Magnitude} &  \multicolumn{3}{c||}{SBO Magnitude} & \multirow{2}{*}{Detector Trip Off} \\\cline{2-7}
        & \multicolumn{1}{c|}{Saturation} & \multicolumn{1}{c|}{Visible} & \multicolumn{1}{c||}{Too Faint} & \multicolumn{1}{c|}{Saturation} & \multicolumn{1}{c|}{Visible} & \multicolumn{1}{c||}{Too Faint} & \\
       \hline
         u & 0.24 & 0.11 & 0.65 & 0.21 & 0.12 & 0.67 & 0.01 \\
         g & 0.48 & 0.09 & 0.43 & 0.45 & 0.10 & 0.45 & 0.02 \\
         r & 0.73 & 0.07 & 0.20 & 0.71 & 0.08 & 0.21 & 0.06 \\
         i & 0.89 & 0.06 & 0.05 & 0.87 & 0.07 & 0.06 & 0.10 \\
         z & 0.98 & 0.02 & 0.0 & 0.97 & 0.02 & 0.01 & 0.14 \\
         Y & 0.99 & 0.01 & 0.0 & 0.99 & 0.01 & 0.0 & 0.20 \\
        \hline
    \end{tabular}
    \caption{Fraction of SNe that will saturate the detectors, will be visible and not saturate the detectors, or will be too faint to detect; with regards to the Peak Magnitude and SBO Magnitude from Figure \ref{fig:ECDF} with a 30 second exposure. The differences between the peak and SBO magnitude is minimal given the assumed 1.4 magnitude difference. We also include the fraction of SNe that will be so bright that we could potentially trip off a detector.
    }
    \label{tab:Sat_sig}
\end{table*}

The LSST TRILEGAL simulation provides the distance and RA/Dec of massive stars in our galaxy. We predict the peak magnitude of galactic CCSNe by combining these locations with SCOTCH explosion models and SFD map extinction. While peak brightness is a good indicator of whether or not we will observe the next CCSN, one of the unique opportunities enabled by multimessenger observatories is the possibility of observing the SBO. SBO is the moment that light first escapes from the CCSN, before the ejecta breakout and nickel decay begins to dominate the light curve. As a rough estimate of the SBO magnitude, we assume that all the explosions will be from a `standard' red super-giant (15 solar mass, 500 solar radii, $10^{51}$ ergs); in that case, \cite{Garnavich2016} shows that SBO will have 25\% of the flux at peak magnitude. 
% The SBO magnitude distributions are the dashed lines in Figures \ref{fig:PDF} and \ref{fig:ECDF}. The intersections of the Rubin limits with the cumulative probability distributions can be found on Table \ref{tab:Sat_sig}

Figures \ref{fig:PDF} and \ref{fig:ECDF} show the probability distribution and cumulative distribution, respectively, of peak and SBO magnitudes for these stars and assumptions. 
For the cumulative plot, we overlay the saturation limits and observing depths for all of the bands assuming the nominal 30 second exposure time. These limits can be shifted to higher magnitudes by increasing the exposure time and lowered by decreasing (up to 1 second). The limit calculations and assumptions can be found in the Appendix. 
Following Figure \ref{fig:ECDF}, the estimated fraction of SNe that fall below the saturation and 5$\sigma$ detection Rubin limits, for each individual band, are shown in Table \ref{tab:Sat_sig}.

\begin{figure*}[!htp]
    \centering
    \includegraphics[width=0.48\textwidth]{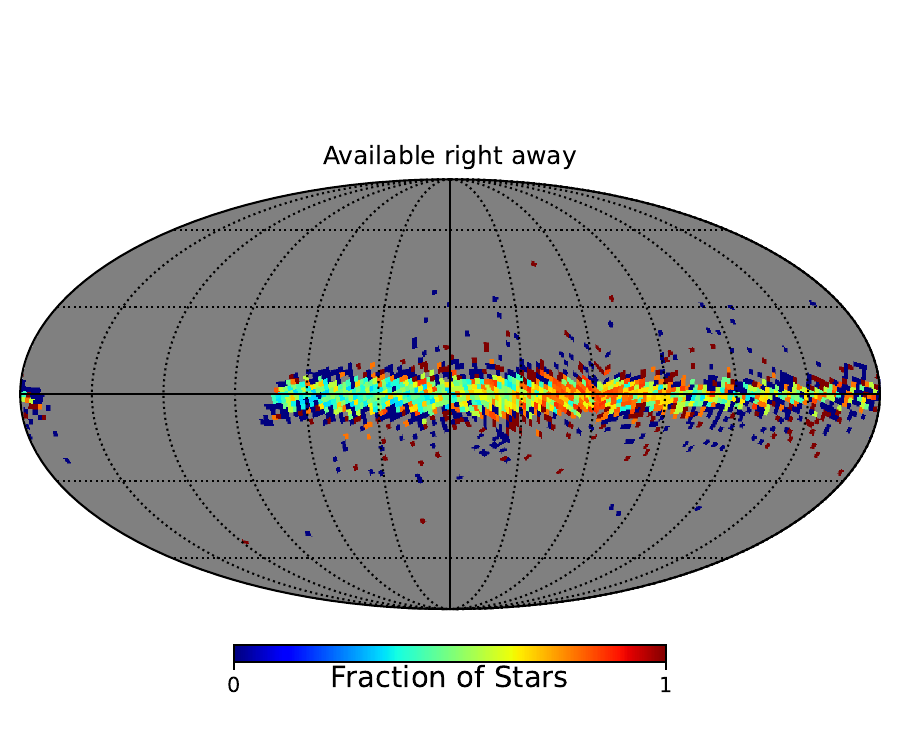}
    \includegraphics[width=0.48\textwidth]{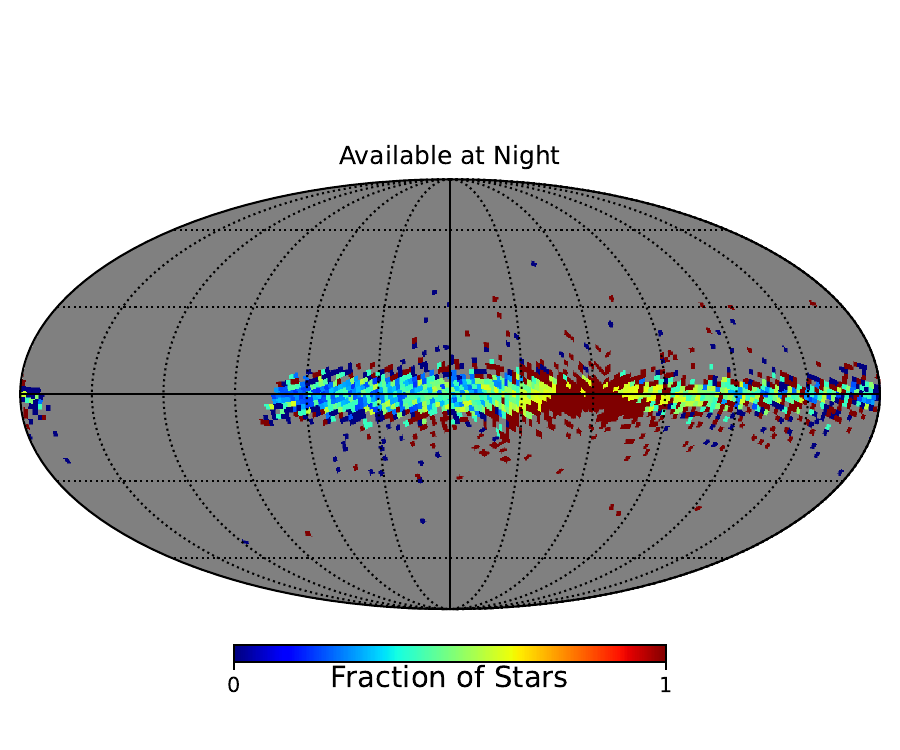}
    \caption{The results of placing a 100,000 CCSN at a random time of the year and random location in our subset of the LSST TRILEGAL catalog. \textbf{Left:} Fraction of stars that explode at night and are available to observe (Available right away). \textbf{Right:} Fraction of stars that explode during the day but are available at night (Available at Night).}
    \label{fig:SNAvailability}
\end{figure*}

\section{Observing Strategy} \label{sec:Discussion}
From the range of magnitudes in Figure \ref{fig:ECDF} we can plan an observing strategy for the next galactic CCSN. Figure \ref{fig:ECDF} and Table \ref{tab:Sat_sig} show that the vast majority of the candidate CCSNe will be observable with Rubin. 
In this section, we characterize other practical concerns (visibility and availability) impacting our ability to catch the SN, or SBO, signal. We then outline recommendations for the follow-up observing strategy.
% Fast slew speed is one of the greatest advantages of using Rubin for follow-up of multi-messenger  galactic CCSN alerts; Rubin can potentially detect the SBO within a few minutes of receiving an alert. The best such alert from a galactic CCSN would come from neutrinos. Neutrinos account for $99\%$ of the CCSN energy budget and can be received seconds to days before the first optical signal. Models suggest that there may be a gravitational wave detection before a star goes CCSN \citep[e.g.][]{Choi2024}, but there has never been such a detection, while neutrinos associated with SN 1987A have been detected \citep{Hirata1987}.

\subsection{Visibility} \label{sec:visibility}
The first factor to characterize Rubin's potential for observing the next galactic CCSN is the likelihood that Rubin is able to image the CCSN. Using random LSST TRILEGAL star locations, we test 100,000 random times over the course of a year and record whether Rubin is able to observe the CCSN. We consider two scenarios: if the CCSN event is at night and is observable, and
% we classify it as ``Available Right Away'', 
if the event is during the day but can be observed at night.
% we label it as ``Available at Night''.
Figure \ref{fig:SNAvailability} shows the smoothed averages of the results for both scenarios as a function of sky position. Rubin will excel at getting the southernmost objects (especially those that are circumpolar at that latitude) and will be less likely able to observe northern CCSNe.

\begin{figure*}[!htp]
    \centering
    \includegraphics[width=0.45\textwidth]{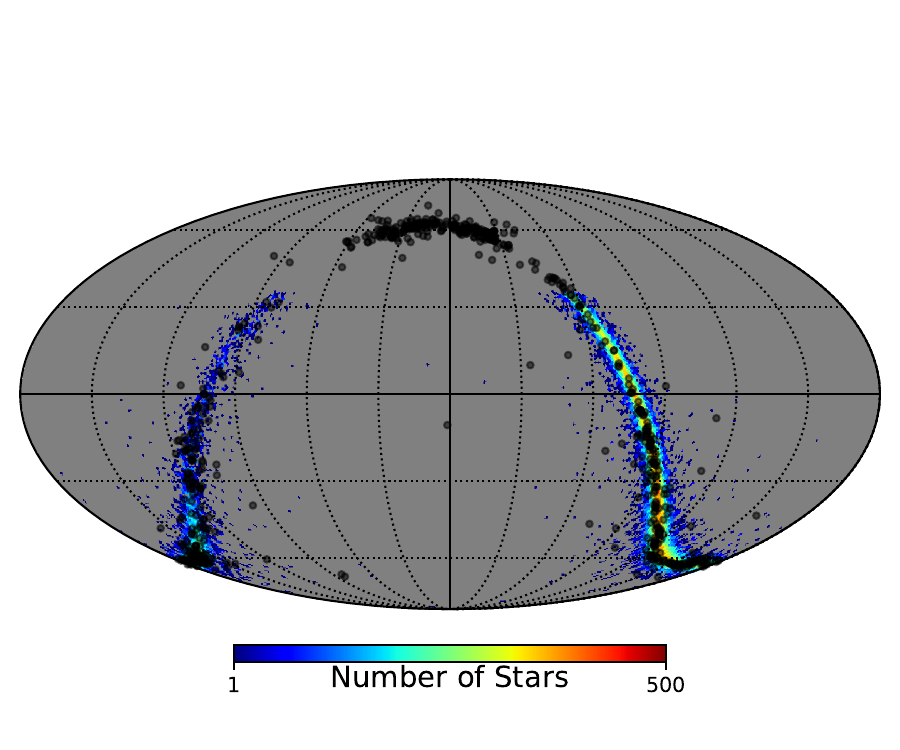}
    \caption{Same as Figure \ref{fig:star_distribution} but overlaid with the locations of the red supergiant candidates from \cite{Healy2024} in black.}
    \label{fig:LSST_Gaia}
\end{figure*}

While Rubin will likely be able to catch the next galactic CCSN, there are a few possible scenarios where it cannot. The prominent one is that the CCSN is too far north in the sky. Using the stellar density assumptions from LSST TRILEGAL, the northern spur of the galaxy should have roughly 1195 CCSN candidates. Adding these candidates to the 39723 that are visible from Rubin shows that LSST will have a 97\% chance of capturing the next galactic CCSN. However, this number might be an optimistic estimate due to the stellar mass density assumptions used to create LSST TRILEGAL. For an observational based and more conservative approach, we use the red supergiant CCSN candidates from \cite{Healy2024} to trace the likelihood of candidates falling into the footprint of LSST. Figure \ref{fig:LSST_Gaia} shows the stellar distribution of the LSST TRILEGAL  catalog overlaid with the red supergiant CCSN candidates of \cite{Healy2024}. This shows that of the 651 candidates, 278 are too far north for Rubin, meaning that only $\sim 57\%$ of these candidates will be visible. While this analysis has its own caveats (e.g.~the catalog only extends to 12.9 kpc), it highlights the difference between the assumptions of LSST TRILEGAL as compared with actual observations.

\subsection{Observability}

Moving forward, we will assume that the CCSN exploded at night and is available to view with Rubin. If it exploded during the day, we will have some time, and maybe even results from other observatories, to optimize observing strategy. For observations of CCSNe exploding at night and available to Rubin, we would most likely be using single exposures and tiling to ensure that we are able to find the galactic CCSN. Depending on the survey progress when the galactic SN goes off, we could possibly have templates of the galactic plane. 

\begin{figure*}[!htp]
    \centering
    \includegraphics[width=0.45\textwidth]{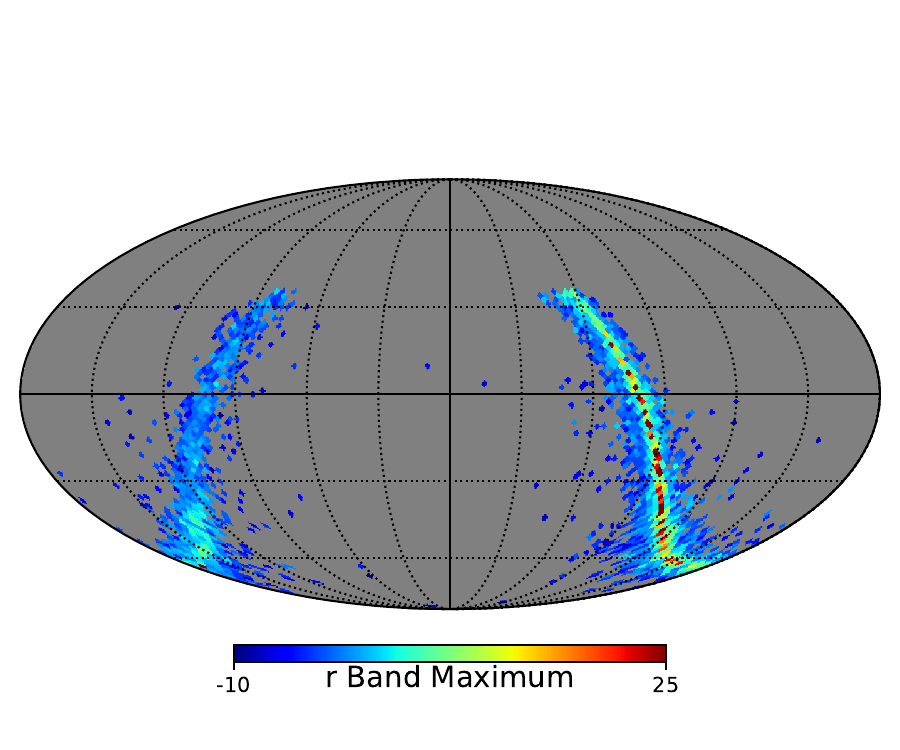}
    \includegraphics[width=0.45\textwidth]{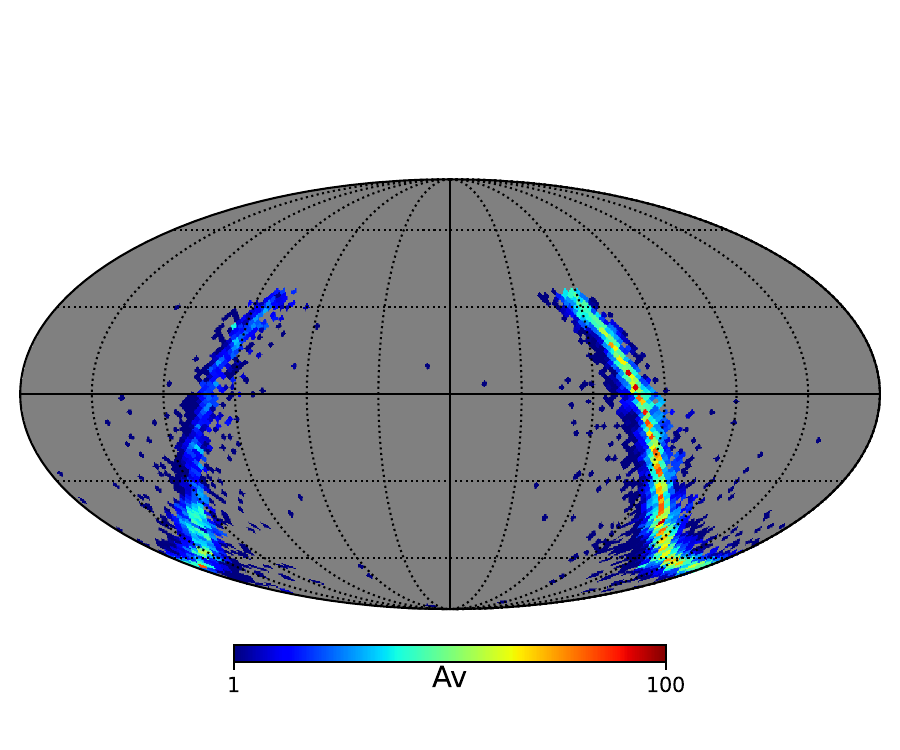}
    \caption{\textbf{Left:} Healpix map showing the distribution of the apparent magnitudes in \textit{r} of the results from Figure \ref{fig:PDF}. This can help us determine the exposure time needed to image the CCSN if given a pointing from Super-K. \textbf{Right:} Same as the left but displaying the results from the $A_{V}$ distribution given by the SFD dust map.}
    \label{fig:Healpix_maps}
\end{figure*}

We can map these peak SN magnitudes spatially to help determine the optimal exposure time as a function of location. The left panel of Figure \ref{fig:Healpix_maps} shows the distribution of apparent magnitudes in \textit{r} band. The large range of magnitudes essentially maps the difference in $A_{V}$ across the Milky Way, shown in the right panel of the same figure. We can utilize $A_{V}$ and dust maps like these to determine the needed exposure times when looking for the CCSN for the various filters.

We also want to see how likely it is that we can see the SBO as a function of location. This should be similar to Figure \ref{fig:Healpix_maps}, but with one major difference: we assume the worst case of SBO occurring only minutes after the neutrino trigger, in a scenario where the time it takes to change the filter (~90-120 seconds) could mean missing the SBO. Similar to what we did in Section \ref{sec:visibility}, we simulate 100,000 explosions at random locations with a random filter in place.
The result follows the same trend as Figure \ref{fig:Healpix_maps}. This is due to the fact that only the $z$ and $y$ filters are able to pierce the center of the galaxy. So, if we are unlucky enough to have a bluer filter in place, and there is a short neutrino trigger to SBO delay, there is a chance that we will miss the SBO. This motivates the need to switch to a redder filter as soon as possible, especially if the neutrino search area is in the center of the Milky Way.

\subsection{Recommendations}

In this paper we have shown that, assuming the SN is observable by Rubin, it is guaranteed to be bright enough to be observed at some point during its light-curve evolution in the two reddest filters, but might not be in the bluer filters.  
A large remaining uncertainty is the time-difference between the explosion and SBO, which determines the lag with which the EM counter-part arrives. 
As the time between the alert and the SBO could range from mere minutes to hours, after the time it takes to slew to the target, a decision will need to be taken as to when and if a filter change should be made to a redder band to increase the probability of catching the SBO. This decision will be heavily influenced by the location of the CCSN and what filter is already in place when it goes off. Optimizing this process will be the focus of future work.

% In order to catch the SBO, the most important response would be to go on sky as soon as possible, with whatever filter is in place at that moment, and return later to replace it with a more favorable red filter. Given that five of the six filters are always available to be loaded in place, we are guaranteed to have either z or Y available. 

There will be many variable objects in the galactic center where the next CCSN is most likely to go off. An advantage in determining which the galactic CCSN is that it gets brighter in time and so sooner or later it will be clear which is the right object. At the same time, variable stars and solar system objects will be an obvious contaminant. Future work will determine the optimal set of selections, tools and pipelines to rapidly carry out this task.
% The stock LSST prompt processing code is  not optimized for a nearly-continuous monitoring of the same patch for a new brightening source and its performance in very crowded fields might be suboptimal. Whether it will nevertheless be the right choice in this context, or whether a simpler, custom processing script is needed is another open problem for future research.
% We therefore suggest to simply use the earlier observations as templates for differencing rather than rely on the latest available system templates which will detect a large number of objects in difference imaging in the first observations.

\begin{figure*}[!t]
    \centering
    \includegraphics[width=0.45\textwidth]{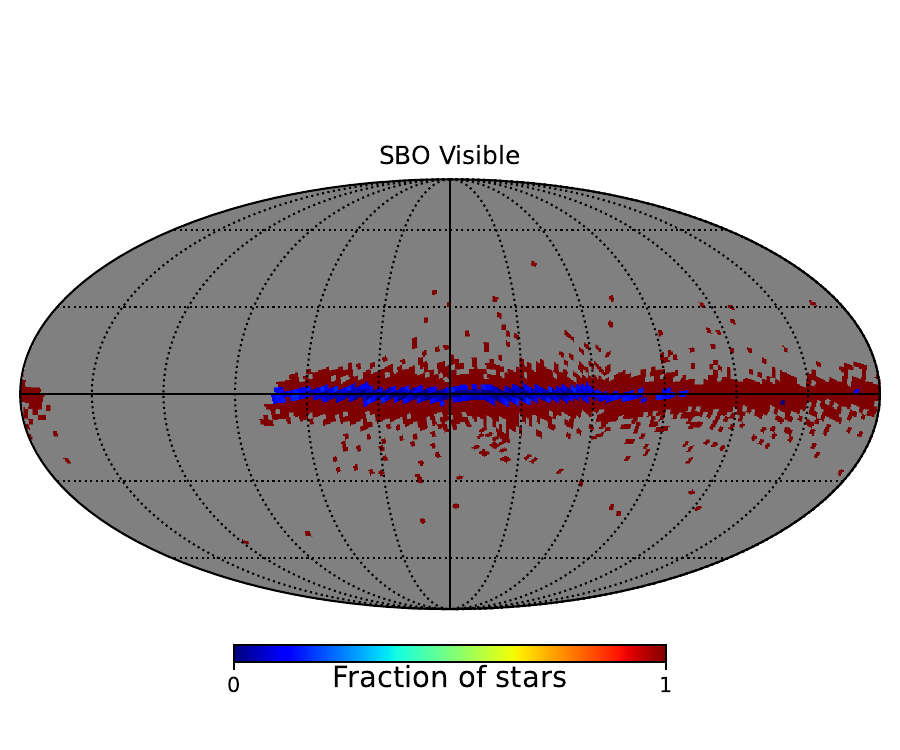}
    \caption{Fraction of stars for which, with a random filter assignment, the SBO is visible.}
    \label{fig:Healpix_SBO}
\end{figure*}

The observing strategy can thus be summarized as follows:
\begin{enumerate}
 \item As soon as triggered, move to the target area and start taking 30s exposures in the current filter. Either utilize existing templates or perform on the fly difference imaging.
 \item In the early times tile over the target region, until a candidate has been found.
 \item After a determined time, change into either z or Y filter, if we are not already using one of those.
 \item When a candidate is found using difference imaging, cross-check its position against known objects in the sky.
 \item Take at least another exposure to ensure the candidate continues to brighten and is consistent with being a SN.
 % and perform quality assurance with a SN expert.
 When convinced, issue an alert with precise localization enabling other observatories to follow up this event.
 \item Rubin should follow the found candidate in the same filter for as long as possible during the first night, to enable cross-calibration of the early light-curve with light-curves taken by other observatories with a later starting time.
\end{enumerate}

One caveat of this strategy is that a galactic SN could be bright enough to potentially trip-off the high voltage back bias (HV bias) of the detectors. Figure \ref{fig:ECDF} shows the latest estimate for the brightness needed to trip-off the HV bias at around -1 mag (private communication) and Table \ref{tab:Sat_sig} shows the percentages.
While even a tripped-off detector gives some pointing information of the SN (each detector covers about 1.7 square arcminutes), the entire readout electric board trips off, disabling 3 of the CCDs for upwards of 10 minutes. One of the clear downsides of opting for a redder filter is that there is a significant increase in risk of a  detector trip-off. There are some mitigations that could be implemented to the above strategy to avoid this. The first would be to use the $A_{V}$ map of the Milky Way (something similar to Figure \ref{fig:Healpix_maps}) to inform our exposure time setting. This would be combined with the filter we currently have in place to help determine the appropriate exposure time. The second mitigation is that we can simply dither while the tripped-off detector is recovering and then shortening the exposure time if the CCSN is unexpectedly bright. These mitigations should help lower the chance of tripping-off a detector while still being able to obtain a precise location of the CCSN.

\section{Conclusions} \label{sec:conclusion}

In this paper, we characterize the capability of Rubin to observe the next galactic CCSN if it explodes during the lifetime of the 10-year survey. To determine how many CCSNe would be visible to Rubin, we combine LSST TRILEGAL, a catalog of simulated stars throughout the galactic plane that are visible with Rubin, and SCOTCH, a catalog of transients including CCSNe. We use CCSN rates derived for our galaxy to find the peak brightness of the SNe, as well as the brightness of the SBO, assuming the SFD dust maps.

Applying Rubin detection limits to these results, we found that almost all possible CCSNe would be visible with a 30 second exposure in redder bands. We also found that with a 30 second exposure, we would catch SBO for close to all of these SNe. 
In order to capture the next CCSN and ideally its SBO, we outline a plan to quickly identify a candidate in order to issue an alert to the wider astronomical community. 

While these results and discussions can serve as a framework for observing the next CCSN, there are still many challenges that will need to be addressed. One of the most important is locating the CCSN in a crowded field, since the galactic plane can have thousands stars per CCD detector. Testing performance of SN recovery with LSST Alert Production in these fields, or creating custom difference imaging pipelines if necessary, is essential and this work is ongoing. Combined with current and future neutrino detectors, soon everything will be in place to quickly locate the next galactic CCSN should it occur during the lifetime of the LSST. This will give us an unprecedented view into both the CCSN process \citep{Boccioli2024} and neutrino astrophysics \citep[see][for review]{Raffelt2025}.

\section*{Acknowledgments}
We thank for Eric Bellm, Kevin Fanning, Saurabh Jha, Sean MacBride, and Giada Pastorelli for useful discussions. This work was supported by Brookhaven National Laboratory LDRD \#24-016.  JB, CAH, AS and PBD acknowledge support from the US Department of Energy under Grant Contract DE-SC0012704. CWW and DS were supported in this work by the Department of Energy, Office of Science, grant DE-SC0010007.

\bibliography{bib}
\bibliographystyle{aasjournal}

\appendix
For calculating the magnitude limits for saturation and $5\sigma$ detections of the different bands, we utilizing equations from a recent LSST technote: https://smtn-002.lsst.io/. 
For the saturation limits, we use Equation 41,

\begin{equation}
    C_{b}=\frac{5,455}{g}10^{0.4(25-m_{o})}\left(\frac{D}{6.5m}\right)^{2}\left(\frac{\Delta t}{30{\rm\ s}}\right)T_{b}
\end{equation}

where $C_{b}$ is the optimal source counts (ADU) and we assume $D=6.5 m$, $g=1.6$ \citep{Roodman2024}, and $T_{b}$ is the throughput integral. 
To calculate the magnitudes we rearrange the above equation to
\begin{equation}
    m_{o}=25-2.5\log_{10}\frac{C_{b}}{\left(\frac{\Delta t}{30{\rm\ s}}\right)\left(\frac{5,455}{g}\right)}\,.
\end{equation}
For the saturation limits, we assume that saturation occurs at 100,000 ADU \citep{Roodman2024}.

To calculate the $5\sigma$ magnitudes, we utilized the following equation from the technote,
\begin{equation}
    m5=C_{m}+dC_{m}+0.50(m_{sky}-21.0)+2.5\log_{10}(0.7/FWHM_{eff})+1.25\log_{10}(expTime/30.0)-k_{atm}(X-1.0)
\end{equation}
\begin{equation}
    dC_{m}=dC^{inf}_{m}-1.25\log_{10}(1+(10^{0.8dC^{inf}_{m}}-1)/Tscale)
\end{equation}
\begin{equation}
    Tscale=expTime/30.0*10.0^{-0.4(m_{sky}-m_{darksky})}
\end{equation}

the values for each individual band can be found in the technote.

\end{document}